RESEARCH ARTICLE  OPEN ACCESS

# Facial Recognition Enabled Smart Door Using Microsoft Face API

Karan Maheshwari [1], Nalini N [2]
SCOPE, VIT University
India

**ABSTRACT**
Privacy and Security are two universal rights and, to ensure that in our daily life we are secure, a lot of research is going on in the field of home security, and IoT is the turning point for the industry, where we connect everyday objects to share data for our betterment. Facial recognition is a well-established process in which the face is detected and identified out of the image. We aim to create a smart door, which secures the gateway on the basis of who we are. In our proof of concept of a smart door we have used a live HD camera on the front side of setup attached to a display monitor connected with the camera to show who is standing in front of the door, also the whole system will be able to give voice outputs by processing text them on the Raspberry Pi ARM processor used and show the answers as output on the screen. We are using a set of electromagnets controlled by the micro controller, which will act as a lock. So a person can open the smart door with the help of facial recognition and at the same time also be able to interact with it. The facial recognition is done by Microsoft face API but our state of the art desktop application operating over Microsoft Visual Studio IDE reduces the computational time by detecting the face out of photo and giving that as the output to Microsoft Face API, which is hosted over Microsoft Azure cloud support.

## I. INTRODUCTION

In today's world of connectivity and smart devices there is an urgent need to modify our existing day to day objects and make them smart, also it is not the era when we can blindly trust the old and conventional security measures, specifically speaking is our door locks. To change and modernize any object we need to eliminate its existing drawbacks and add extra functionality.

The major drawbacks in a common door lock is that anyone can open a conventional door lock by duplicating or stealing the key and its simply impossible if we want our friends and family to enter our house, without being actually present over there. Thus why not just eliminate these problems. So, to simply convert this normal door lock into a smart lock, which can open the door whenever we turn up in front of the gate or when we want it to open up for someone else without being physically present, we need to modify the door. So an era has come where devices can interact with its users and at the same time ensure of their safety and keep improvising themselves.

The major concepts used to design and model this access control system is advanced knowledge of micro controllers and interfaces, as the Raspberry Pi computing device is used and interfaced with different drivers along with application development to develop a desktop application. Live high quality HD Camera is connected with the display using the same processor to provide the functionalities mentioned above. By capturing the photo and processing through the app which detects face out of the image and sends it over the Microsoft FACE API for recognition. In addition to this different IoT protocols and APIs have been used to make the device smart. An intensive study of opencv platform and its inbuilt libraries has been conducted to generate a code, which does correct and reliable facial recognition with new and efficient use of hardware. As the world is progressing people are scared about the safety of their possessions, information and themselves. With the model of Smart Door a profound impact is expected on the security industry and it is somewhat anticipated as the time has come to make all daily life objects interconnected and interactive. This model will be a major contribution to the field of Home Security.

## II. RELATED LITERATURE AND COMPARISON WITH EXISTING MODELS

Since 2010 the industry has seen a dawn of work being done in fields of Artificial Intelligence, Machine Learning, Neural Networks, IOT, Big Data Analytics all with a common goal to make things easier, self supervising and to interconnect all kinds of devices by making everyday objects interconnected and interoperable.

A need has been felt in the field of digitalizing conventional security tools and thus a lot of work has been modelled on making daily life locks smart by introducing locks movable with the help of stepper motors and/or adding a digital number pad to take inputs from user or adding Infra Red or Bluetooth modules to operate these devices.

An intensive study of literatures implementing Smart Locks had been done and literatures implementing Door Locks with the help of GSM phones [1] and Stepper motors have been studied. Also literatures regarding smart display have been thoroughly reviewed [2]. The fault in existing models is complexity of system and unnecessarily relying on extra components. Our model is unique with its one of a kind combination of functionalities offered and the simplicity of the model. A major difference is in the overhead reduction by the application as it detects the face out of the image and sends it directly to application program interfaced with our application, which has not been provided in any existing model and the efficient use of solenoids, which also eliminates use of stepper motors. So, we have avoided the use of unnecessary components like stepper motors and drivers as done in existing models and also we have given newer and unprecedented features of facial recognition as an access point control system with a combination of relay module with solenoid to open the gate and unique and interactive User Interface. Also rather than using a low quality Raspberry Pi Interfaced Camera we have used USB attachable HD WebCam to do efficient and reliable facial recognition.

The objectives of the proposed work is to implement a working model of a smart door and to give solutions to the problem faced by people in day to day incidents of burglary or loosing the key and also to promote and ignite the work being done on IOT systems and implementing it with the help of key research areas of Neural Networks and IoT APIs and protocols.

This model is allowing people to add more functionality to it and thus induce more research work in the field of AI, Machine Learning, IoT and lot more.

## III. PROPOSED METHODOLOGY

In order to implement the smart door model we need a list of materials which is briefly mentioned below:

**Hardware**
Raspberry Pi 3 Model B+
Microsoft HD Live WebCam
Solenoid
Display Monitor
Audio Speaker
Generic Mouse
Generic Keyboard
Relay Module 1 way 5V
Breadboard, Push Buttons and Connecting Wires

**Software**

Microsoft Visual Studio 2015
Microsoft Azure subscription
Microsoft FACE API
WINDOWS 10 IOT
WINDOWS 10 IOT Dashboard Manager
Application with UI deployed in XAML and backend in C#

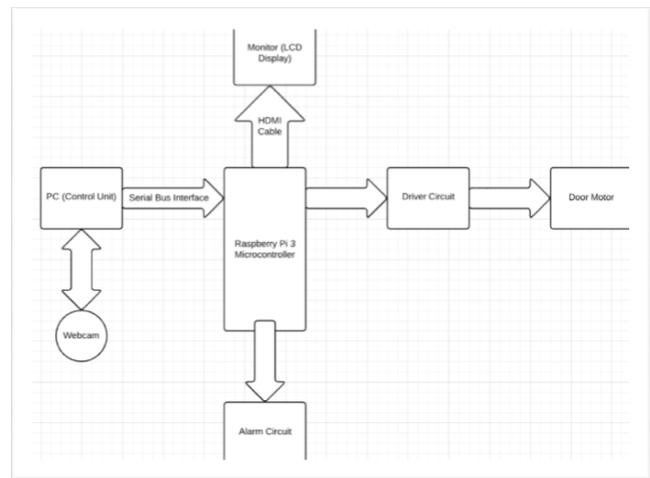

Fig. 1 Data Flow Diagram

To start with first of all we need a Raspberry Pi set up with Windows 10 IOT edition, then interfacing the ARM processor with the display, the display is attached to a camera interfaced with the processor to provide input of who is accessing the door and to capture the image to apply facial recognition computing via Raspberry Pi. The processor is attached with a two-way relay module, which is in turn attached with a solenoid. Processor is attached with push button, which has the functionality of doorbell.

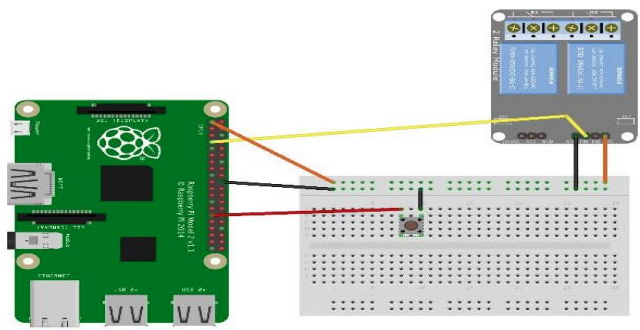

Fig.2 Circuit Diagram

We can segregate the whole system on basis of two major functionalities:
1) Accessing door on basis of recognition: On pressing the door bell the HD camera captures the photo and then the application developed detcts the face out of the image and crops it out and sends it over to the Microsoft Face API interfaced to the application through Microsoft Azure cloud set up. The face is identified and recognized from a pre-saved database of facial images on cloud. If the face is matched User gets an audio of "welcome USER NAME" by and the processor controlling the relay module opens up the door, which can be seen by movement of solenoid.
2) Adding users to a database to be recognized from: To add users we need to click on add user button and on doing this, the camera attached captures the image of user sitting in front and asks for name as input, and adds it to the database of images over the cloud from which the face will be recognized.

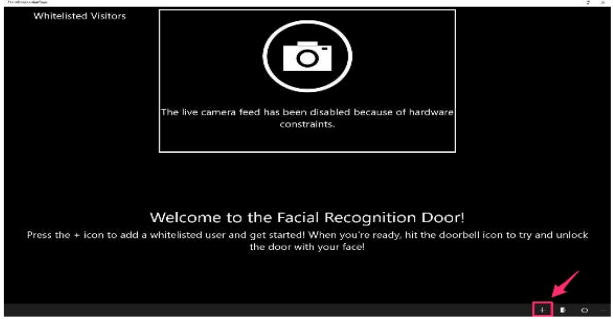

Fig.3 Adding User Functionality

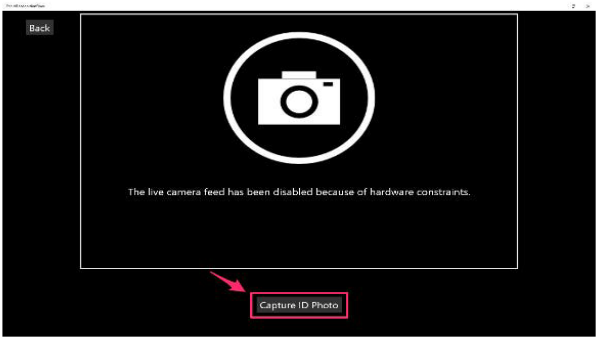

Fig.4 Adding User Functionality

Database is maintained for registered users on the cloud support and temporary database storage for guest users.

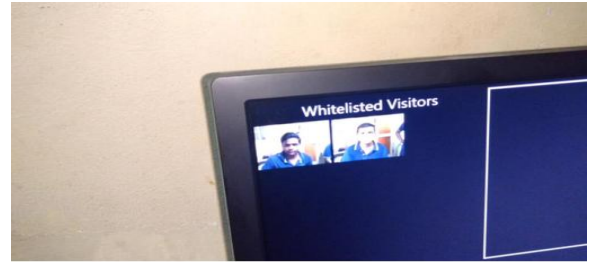

Fig.5 Whitelisted Users

## IV. STEPS TO DEPLOY THE SYSTEM

The device will function with an application and will have an interactive UI, which will let the application call the APIs to access the hardware and do the desired functions and also display the result. The application's front end is deployed on Microsoft Visual Studio and the API(s) are called through it.

*STEPS:*

1) Connect the Raspberry Pi with a power source and install the OS (Windows 10 IOT) on the micro SD card attached to it.
2) Connect the Raspberry Pi ARM controller with the peripherals (display, keyboard, webcam and mouse).
3) On another parent device install Windows 10 IOT Dashboard and configure the RPI processor with it (can be done wirelessly also).
4) Download Microsoft Visual Studio on the parent device and deploy the code generated in the high level language of XAML and C#
5) The deployed app should be interfaced with the API over Microsoft Azure.

As the facial recognition works on the principle of neural networks it requires high computing resources to compute the result, which can be done by GPU support availed through application deployment on Azure.

## V. CONCLUSIONS

In this work, automatic door access system by using face recognition and detected is presented. Automatic face recognition is done by Neural Networks. Raspberry Pi controller controls the door access after successful output from the PC. Immediate responses from the door and monitor are observed. The door remains open for indefinite time and this is not suitable for real time. So, appropriate time should be set in real time environment. This system can be used in many places where need of security is maximum and security cannot be compromised

## VI. LIMITATIONS

The algorithms used in this work are implemented in neural networks and neural networks has few limitations in real time. These are enlisted as:
- Process is slow.
- Result is not so accurate
- The methodology is complex.

Also the time after successful detection of face is indefinite. This time should be set according to user needs. This system has only one admin and this may propose a problem if that person is not available to add any other user in case of emergency.

## VII. FUTURE WORK

- If a blacklisted person tries to open the door, the system will send a message to the admin using GSM module regarding the same.
- A real time speaking assistant can be deployed to make the system more user friendly and efficient.
- Database can be linked to cloud in case of power failures and data loss.
- Highly secure protocols such as TLS can be deployed to ensure there is no security breach.

There is a lot of scope today in the field of home automation and thus we can later enable GSM and DTMF module in this giving more functionality. Also on larger scale rather than using ARM processor we can use x-86 processors to enable live feed.

## ACKNOWLEDGMENT

Both the authors wish to acknowledge Microsoft for developing pay per use softwares and APIs.